\documentclass[11pt]{article}

\usepackage[margin=3cm]{geometry}
\usepackage{caption}
\usepackage{subcaption}
\usepackage{graphicx}%
\usepackage{multirow}%
\usepackage{amsmath,amssymb,amsfonts}%
\usepackage{amsthm}%
\usepackage{mathrsfs}%
\usepackage{latexsym}%
\usepackage{mathtools}%
\usepackage[title]{appendix}%
\usepackage{xcolor}%
\usepackage{color,cancel}
\usepackage{textcomp}%
\usepackage{manyfoot}%
\usepackage[utf8]{inputenc}
\usepackage{listings}%

\usepackage{tocloft}
\usepackage{authblk}
\usepackage{lipsum}
\usepackage{wrapfig}
\usepackage{enumerate}
\usepackage{soul}
\usepackage{chemarrow}

\usepackage{booktabs}%
\usepackage{multirow}
\usepackage[math]{cellspace}
\usepackage{algorithm}%
\usepackage{algorithmicx}%
\usepackage{algpseudocode}%
\usepackage{listings}%
\usepackage{units}
\usepackage{verbatim}
\usepackage{fancyvrb}
\fvset{fontsize=\normalsize}

\usepackage[T1]{fontenc}
\usepackage[sc]{mathpazo}
\setkeys{Gin}{width=\linewidth,totalheight=\textheight,keepaspectratio}

\usepackage{hyperref}
\hypersetup{
	hidelinks = true
}

\usepackage{amsmath}
\usepackage{bm}
\makeatletter
\renewcommand*\env@matrix[1][*\c@MaxMatrixCols c]{%
  \hskip -\arraycolsep
  \let\@ifnextchar\new@ifnextchar
  \array{#1}}
\makeatother
\title{Nucleation feedback can drive establishment and maintenance of biased microtubule polarity in neurites}

\author[1]{Hannah G. Scanlon} 
\author[2]{Gibarni Mahata}
\author[3]{Anna C. Nelson}
\author[4]{Scott A. McKinley}
\author[2]{Melissa M. Rolls}
\author[1,5]{Maria-Veronica Ciocanel\thanks{\url{veronica.ciocanel@duke.edu}}}

\affil[1]{Department of Mathematics, Duke University, Durham, NC, 27710, USA}

\affil[2]{Department of Biochemistry and Molecular Biology, Pennsylvania State University, University Park, PA, 16802, USA}

\affil[3]{Department of Mathematics \& Statistics,  University of New Mexico, Albuquerque, NM, 87131, USA}

\affil[4]{Department of Mathematics, Tulane University, New Orleans, LA, 70118, USA}

\affil[5]{Department of Biology, Duke University, Durham, NC, 27710, USA}

\begin{document}
\maketitle
\begin{abstract}
The microtubule cytoskeleton is comprised of dynamic, polarized filaments that facilitate transport within the cell. Polarized microtubule arrays are key to facilitating cargo transport in long cells such as neurons. Microtubules also undergo dynamic instability, where the plus and minus ends of the filaments switch between growth and shrinking phases, leading to frequent microtubule turnover. 
Although microtubules often completely disassemble and new filaments nucleate, microtubule arrays have been observed to both maintain their biased orientation throughout the cell lifetime and to rearrange their polarity as an adaptive response to injury. 
Motivated by cytoskeleton organization in neurites, we propose a spatially-explicit stochastic model of microtubule arrays and investigate how nucleation of new filaments could generate biased polarity in a simple linear domain. Using a continuous-time Markov chain model of microtubule growth dynamics, we model and parameterize two experimentally-validated nucleation mechanisms:
nucleation feedback, where the direction of filament growth depends on existing microtubule content, and a checkpoint mechanism, where microtubules that nucleate in a direction opposite to the majority experience frequent catastrophe. When incorporating these validated mechanisms into the spatial model, we find that nucleation feedback is sufficient to establish biased polarity in neurites of different lengths, and that the emergence and maintenance of biased polarity is relatively stable in spite of stochastic fluctuations. This work provides a framework to study the relationship between microtubule nucleation and polarity, and could extend to give insights into mechanisms that drive the formation of polarized filament arrays in other biological settings. 
\end{abstract}

\begin{footnotesize}
\end{footnotesize}

\section{Introduction}
\label{intro}

Microtubules (MTs) are polar protein filaments with distinct plus and minus ends, which provide structure to cells as part of the cytoskeleton. A key role of microtubules is to facilitate protein transport and sort cellular content, as motor proteins recognize microtubule polarity and walk towards either their plus or minus end, carrying cellular cargoes to specific destinations. Very long cells, such as neurons, have a crucial need for directional transport, as cellular cargoes must often travel long distances. However, individual microtubules do not span the full length of neuronal cells. For example, lower motor neurons in our legs may be up to a meter long, but individual microtubules in neurons have been measured in the range of tens of microns \cite{Chalfie1979Organization, Rolls2021Microtubule, Yogev2016MTOrganization, Yu1994MTChanges}. Given this difference in spatial scales, many differentiated cells have highly polarized microtubule arrays that span the necessary spatial domain with overlapping microtubules that continuously cover their neurites \cite{Bartolini2006NoncentrosomalArray}. Furthermore, microtubules experience a phenomenon called dynamic instability, in which both ends switch between periods of growth and shrinking \cite{Mitchison1984Dynamic}. Given these dynamics, microtubule content regularly experiences turnover in living cells, but the polarized organization of the MT array is often maintained for the lifetime of the cell. 

We previously used mathematical modeling to study microtubule turnover in neuronal dendrites and predicted plausible length-limiting mechanisms \cite{2024NelsonMinimalMech}. Recently, we expanded this approach to incorporate filament nucleation and predict tubulin usage and microtubule length distributions under different nucleation regimes \cite{nelson2025emergent}. However, many questions remain about the spatial regulation of microtubule nucleation and how this process contributes to polarity regulation. For example, in \textit{Drosophila} sensory neurons, microtubules can be nucleated with different orientations, but only recently have studies suggested mechanisms that prevent dendritic nucleation from disrupting polarity \cite{Feng2021ParallelPolymerization}. Here, we focus on mechanisms that enable dynamic microtubules to maintain a highly polarized, noncentrosomal array in long linear domains such as dendrites and axons. We are particularly interested in how mechanisms that control microtubule nucleation can contribute to biased polarity in these domains. Using a stochastic modeling framework and experimental validation, we find that a nucleation feedback mechanism is sufficient to drive biased microtubule polarity in neurites, and we predict that different biased polarities can emerge.

\subsection{Variety and adaptability of polarized microtubule arrays}

The question of microtubule polarity is particularly interesting in neurons since neuronal microtubules have different polarized arrangements in axons and dendrites \cite{Rolls2015,Baas2011,2021WeinerHausNucleation,iwanski2023cellular}. Across all species studied, axonal microtubules are consistently organized with plus-end-out (PEO) polarity, where plus ends are directed away from the cell body \cite{Stone2008,Burton1981,Harterink2018,Heidemann1981,Lee2017,Stepanova2003,Goodwin2012, Thyagarajan2022}. In contrast, dendrites have a significant portion of minus-end-out (MEO) microtubules, with minus ends directed away from the cell body, and mature dendrites in invertebrates show an almost uniform MEO polarity \cite{Stone2008,Harterink2018,Rolls2022}. 

Moreover, after axonal injury, microtubules are capable of completely reorganizing their polarity to facilitate axonal regeneration  \cite{Stone2010}. For example, in \textit{Drosophila} dendrites, MTs in a selected dendrite have been shown to reverse their polar arrangement following axon injury, in order to become an axon with PEO MTs \cite{Stone2010}. This ability to properly reorient microtubules is crucial to regeneration and failure to reorganize has been shown to block regeneration~\cite{Mattie2010MotorSteering}. 
We are thus interested in identifying and validating polarity regulation mechanisms which can produce microtubule arrays of distinct orientations in different cellular regions (such as axons and dendrites). In addition, these mechanisms must be flexible, so that neurons can respond dynamically in the case of injury. Uncovering such mechanisms is crucial in understanding both healthy neuronal function and resilience.

\subsection{Proposed mechanisms of microtubule polarity regulation}

Several studies have focused on how nucleation of new microtubules interacts with or contributes to polarity regulation. One possibility is that nucleation is completely uncoupled from polarity control. One of the first biological models for organization of microtubule polarity in neurons took this approach. In this microtubule sliding model, microtubules are nucleated at the centrosome in the middle of the neuron and then motor proteins slide them in the correct orientation into the axon \cite{Guha2021MTSliding}. However, it is now clear that microtubules can be generated by nucleation throughout long cellular extensions, such as neuronal axons and dendrites \cite{Vinopal2025AcentrosomalNucleation, 2021WeinerHausNucleation}. 

In long neurites, nucleation is likely to be linked with MT polarity. Two biological models have been proposed for how tightly-controlled microtubule nucleation could contribute to the generation of polarized microtubule arrays in long cellular processes. \textit{C. elegans} sensory neurons provide an example of the first model, where microtubules have been tracked as a dendrite emerges from the cell body. As the new dendrite extends, a microtubule organizing center (MTOC) located near the dendrite tip seeds the dendrite with new microtubules nucleated with their plus ends pointing towards the cell body \cite{Liang2020DendriteGrowthCone}. This model uses nucleation from a point source to establish MEO polarity, but does not address how this polarity would be maintained over time through microtubule turnover or how MT arrays with different polarity distributions could be established.

A second model has focused on how nucleation might be used to maintain polarity in long cellular processes. The HAUS or augmin complex can recruit nucleation sites to the sides of pre-existing microtubules and typically nucleates new microtubules with similar orientation to the pre-existing microtubule \cite{Kamasaki2013Augmin, Petry2013BranchingAugmin}. Consistent with this model, studies in axons have shown microtubule polarity becomes more mixed when HAUS subunits are reduced \cite{Cunha2018Haus, Sanchez2016Augmin}.  This model of nucleation control could account for polarity maintenance but likely only applies to a subset of scenarios and its stability over time has not been explored. 

Here, we propose and analyze a mathematical model of microtubule polarity and nucleation inspired by and parameterized using \textit{Drosophila} sensory neuron data to explore two, more flexible mechanisms that have been shown to influence polarity of newly-nucleated microtubules. First, in mature \textit{Drosophila} dendrites, nucleation has been found to be biased towards the cell body, aligning with the observed MEO polarity distribution \cite{Yalgin2015,Weiner2020,Feng2021ParallelPolymerization}. Previous work has not identified a mechanism for this bias, or whether this nucleation bias is dependent on pre-existing microtubules. Second, microtubules in this system undergo a checkpoint mechanism which is dependent on pre-existing MT content. After nucleation, microtubules that were nucleated with orientation opposite to the majority undergo frequent catastrophe, while those generated with orientation parallel to the majority have their plus-end growth promoted \cite{Feng2021ParallelPolymerization}. We investigate these two flexible, nucleation-based mechanisms to understand their function and determine if they can enable robust polarization of microtubule arrays, which may be applicable to a variety of cellular processes.

\subsection{Mathematical modeling of microtubule dynamics and organization}

The mechanisms influencing microtubule organization and polarity in neuronal structures are complex and their interactions or redundancies are not well understood. To overcome this challenge, we propose a mathematical modeling framework that focuses on a simple scenario -- a linear region of a neurite with periodically-spaced nucleation locations -- and explores the interplay between microtubule growth, nucleation, and polarity. 

Mathematical models for different levels of microtubule behavior have been proposed in previous studies, from modeling detailed behavior of individual microtubule ends to more global analyses of the mitotic spindle. However, fewer studies have focused on modeling and validating the control mechanisms for microtubule growth and spatial organization in the cell. Previous spatial mathematical models of microtubule organization have focused on polarized arrays of microtubules that are organized and nucleated by a central MTOC. These studies considered stochastic and agent-based models of chromosome alignment \cite{gardner2005tension}, assembly of the mitotic spindle \cite{paul2009computer}, assembly of an integral Golgi complex \cite{vinogradova2012concerted}, and centration and rotation of the pronuclear complex \cite{coffman2016stronger}. Other studies used computational approaches to understand how microtubule nucleation is regulated during cell division \cite{gregoretti2006insights}. All these studies model microtubules that only undergo dynamic instability at their plus ends, and they are typically interested in the movements of biological structures (such as kinetochores, the Golgi complex, etc.) that interact with microtubules. Many of these works choose parameters that ensure qualitative rather than quantitative fits to experimental data, and they incorporate mechanisms such as spatially-biased catastrophe or growth \cite{gardner2005tension,paul2009computer,gregoretti2006insights} or random microtubule nucleation \cite{vinogradova2012concerted,coffman2016stronger}.In addition, models of microtubule networks in neurons tend to focus on the transport and segregation of cargo proteins such as neurofilaments \cite{xue2015stochastic,lai2018stochastic,ciocanel2020mechanism}. Computational models have also been used to study different mechanisms of microtubule organization in plant cells, where spatial effects like microtubule collision play a key role in establishing correct microtubule orientation \cite{allard2010mechanisms,ambrose2011clasp}.  

Our goal is to understand the relationship between MT nucleation and polarity in neurons, so we build a spatially-explicit mathematical model of polarized arrays of microtubules that are organized independently of an MTOC, as is seen in neurons. To accurately represent microtubule growth and shrinking dynamics, we use our previously-developed stochastic model of microtubule dynamic instability, which incorporates biological measurements at plus and minus ends of microtubules and accounts for minimal mechanisms of microtubule length regulation \cite{2024NelsonMinimalMech}. Building on this model of microtubule turnover, we develop a spatially-explicit model of microtubule polarity and organization in a linear neurite domain with periodically-spaced nucleation locations. This modeling framework allows us to investigate nucleation-based mechanisms, such as the nucleation feedback and the checkpoint mechanisms which were identified in \textit{Drosophila} dendrites \cite{Feng2021ParallelPolymerization}, and their interplay with neuronal polarity. Unlike prior works, we model the dynamic instability of noncentrosomal microtubules at both their plus and minus ends, we parameterize the hypothesized mechanisms using experimental data, and we are particularly interested in nucleation-based feedback mechanisms that influence microtubule polarity through space and time. By modeling a simplification of the typically branched structure of a dendrite with a linear domain, we are able to fully characterize the roles of these identified mechanisms while still employing a realistic, representative domain.

\subsection{Overview}

In Section~\ref{sec:nucleation_bio_data}, we present new experimental results that shed light on a mechanism that regulates biased microtubule nucleation in \textit{Drosophila} neuron dendrites. In Section~\ref{sec:checkpoint_mechanism}, we derive and parameterize a checkpoint mechanism that validates experimental observations. In Section~\ref{sec:math_model}, we describe our mathematical modeling framework. We describe our construction of a spatially-explicit microtubule model in Section~\ref{sec:spatial_model}. In the rest of Section~\ref{sec:math_model}, we summarize our prior stochastic model of microtubule growth and shrinking dynamics  and describe the implementation of nucleation-based mechanisms of polarity regulation suggested by experiments. In Section~\ref{sec:results}, we demonstrate the impact of nucleation mechanisms on polar microtubule organization, and we show model predictions of the emergence of stable, biased MT polarity in linear neurites. We discuss our results in Section~\ref{sec:discussion}.

\section{Nucleation feedback: experimental methods and results}
\label{sec:nucleation_bio_data}

\begin{figure}[h!]
    \centering
    \includegraphics[width=\linewidth]{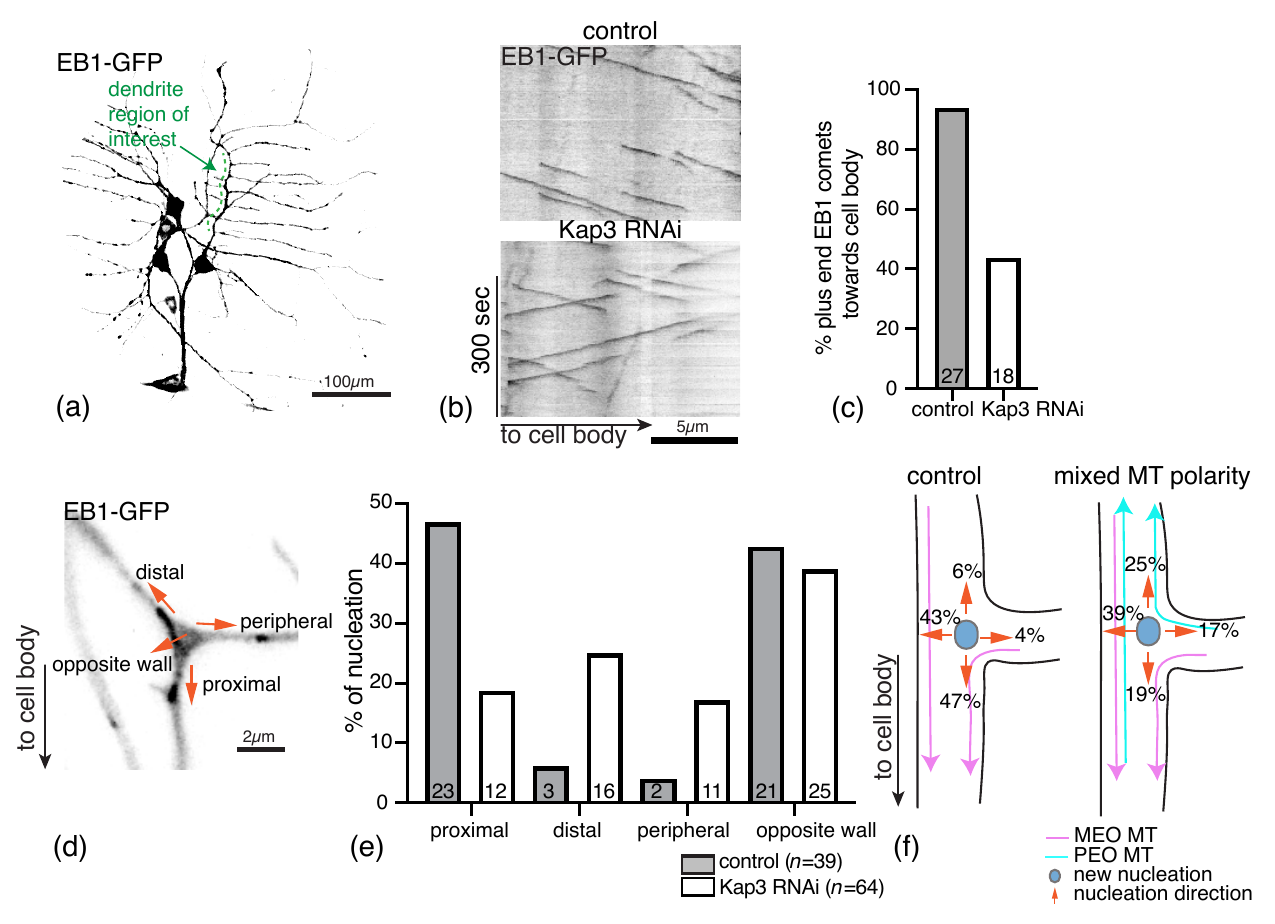}
    \caption{Nucleation direction of new microtubules is dependent on pre-existing microtubule orientation. (a) Drosophila Class I sensory neuron ddaE. (b) Example kymograph of EB1-GFP in control and Kap3 RNAi conditions. (c) Percentage of MT plus ends growing towards the cell body. Numbers on the bars represent the number of cells analyzed and $p<0.00001$ using Pearson's chi-squared test. (d) Dendrite branch point in ddaE neuron with labeled nucleation directions: proximal, distal, peripheral and opposite wall. (e) Percentage of nucleation events in each direction in control and Kap3 RNAi conditions. Numbers on the bars represent the number of nucleation events identified and $p<0.00063$ using Pearson's chi-squared test. (f) Schematics showing the influence of pre-existing microtubules on new nucleation directions. }
    \label{fig:biodata}
\end{figure}

Several previous studies have noted that newly-nucleated microtubules in dendrites have a bias towards the cell body \cite{Feng2021ParallelPolymerization, Yalgin2015}, but what leads to this bias is unclear. To investigate whether pre-existing microtubules influence the nucleation direction of microtubules, we monitored the growth directions of newly-nucleated microtubules in \textit{Drosophila} ddaE dendrites, shown in Figure \ref{fig:biodata}a, and compared the proportions in each direction between dendrites with normal, MEO polarity and dendrites with mixed polarity. We hypothesize that the nucleation direction of new microtubules is dependent on the surrounding microtubule polarity.

\subsection{Drosophila strains}
To study microtubule plus-end growth in Drosophila sensory neurons, a tester line carrying 221-Gal4, UAS-EB1-GFP, nlsBFP-UAS-Dicer2 was used. As a control, we targeted $\gamma$-tubulin37C, which is maternally expressed and does not seem to have a role in somatic cells \cite{Wiese2008}. RNAi lines were obtained from the Vienna Drosophila Resource Center (VDRC): $\gamma$-tubulin37C (VDRC 19130) and Kap3 (VDRC 45400). Flies were grown in standard fly media containing yeast, cornmeal, sucrose, dextrose, and agar and maintained at 25°C. Crosses were set up in vials, and third-instar larvae were selected on the third day for experiments.

\subsection{Live imaging of sensory neurons in Drosophila larvae}

Three-day-old larvae were mounted between a glass slide and a coverslip and secured with tape to immobilize the larvae. Live imaging of whole larvae was performed using a Zeiss LSM800 confocal microscope with a 63x oil objective (NA 1.4). Plus-end growth and nucleation events were recorded in ddaE neurons as five-minute videos that included 150 frames. All videos were analyzed using Fiji (doi:10.1038/nmeth.2019) and kymographs were generated using KymographBuilder (https://github.com/fiji/KymographBuilder) to capture microtubule polarity and nucleation events across time.

\subsection{Generating mixed polarity dendrites}

To generate dendrites with mixed polarity, we used Kap3 RNAi to disrupt microtubule steering at branch points. Kap3 is one of the three subunits of kinesin-2 that is required to steer growing microtubules along existing ones in Drosophila neurons \cite{Mattie2010}. The Kap3 RNAi hairpins were expressed in Class I sensory neurons including the ddaE cell, where we have previously performed mechanistic characterization of microtubule behavior, using the Gal4-UAS system \cite{Brand1993}. They were expressed together with EB1-GFP, which can be used to monitor the behavior of growing microtubule ends in neurons \cite{Stepanova2003}. 

\subsection{Data quantification and visualization}
For microtubule polarity, EB1-labeled plus end growth was quantified in the primary dendritic comb beginning at the first branch point and extending 10-20 $\mu$m distally (Figure~\ref{fig:biodata}a). An EB1-GFP comet was analyzed if it was in focus for at least three consecutive frames. The microtubule polarity was determined by calculating the ratio of plus ends growing toward the cell body divided by the total number of observed comets.

To identify nucleation events, new EB1 comets were tracked in the branch points of the primary dendritic comb where we have previously characterized nucleation \cite{Nguyen2014gamma_polarity}. A nucleation event was defined as the emergence of a new EB1 comet in a region where no previous comets were observed, provided the comet remained in focus for at least three frames. To ensure the accuracy of tracking nucleation events, the region around the comet was confirmed to be in focus before and after the nucleation event. The initial nucleation growth direction was recorded. We categorized the nucleation directions as proximal, distal, peripheral and opposite wall defined relative to the cell body (Figure~\ref{fig:biodata}d). Any obvious severing, rescue events or nucleation outside the branch points were excluded from the analysis. The number of nucleation events in each direction was quantified and divided by the total number of nucleation events to determine the frequency of nucleation in each direction.

\subsection{Experimental results}
As expected, knockdown of Kap3 disrupted the uniform polarity of dendritic microtubules, resulting in mixed polarity. In control ddaE neurons, 94\% of plus-ended EB1-GFP comets moved towards the cell body, whereas in Kap3 RNAi, only 44\% of plus ended EB1-GFP comets showed this directionality (Figure \ref{fig:biodata}b, c). In these distinct polarity backgrounds, we analyze the nucleation direction of new EB1-GFP comets at branch points. In the control ddaE neurons, new EB1-GFP comets demonstrate a nucleation bias either towards the proximal direction (47\%) or the opposite wall (43\%). However, in Kap3 RNAi nucleation events are less biased towards the proximal direction with only 19\% of the new comets. Instead, we observe that nucleation increases significantly in the peripheral and distal directions in Kap3 RNAi dendrites compared to the control (Figure \ref{fig:biodata}e). Thus, we conclude that pre-existing microtubules instruct the directionality of new nucleation events (Figure \ref{fig:biodata}f). This finding suggests that, in our mathematical model, the direction of new microtubule nucleations should be dependent on local pre-existing microtubules through a nucleation feedback mechanism. 

\section{Checkpoint mechanism derivation}\label{sec:checkpoint_mechanism}

In Section~\ref{sec:nucleation_bio_data}, we consider control of the direction of new microtubule nucleations. However, the biological findings in \cite{Feng2021ParallelPolymerization} suggest that, after this step, there is an additional quality control or checkpoint mechanism that controls the fate of these new microtubules. Specifically, the study finds that nucleations with the same orientation as pre-existing microtubules are more likely to grow at least $2 ~\mu$m past the branch point at which they were nucleated than those with the opposite orientation. \cite{Feng2021ParallelPolymerization} measures checkpoint success rates of PEO and MEO MT nucleations in two different dendritic polarity environments. These data are depicted in Figure~\ref{fig:checkpoint_construction}a and recorded in Table~\ref{tab:model_parameters}. However, it remains challenging to determine how this checkpoint mechanism functions purely based on these biological observations. 

In this section, we hypothesize realistic designs of the checkpoint mechanism and investigate whether they are consistent with the biological observations in \cite{Feng2021ParallelPolymerization} using theoretical probability calculations. 
We first calculate the baseline probability that a new nucleation grows at least $2~\mu$m past its nucleation site, in the absence of any mechanism, in Section~\ref{sec:growth_baseline}. In Section~\ref{sec:environment_definitions}, we define the two polarity environments tracked in experiments and relevant to our calculations. In Section~\ref{sec:simple_checkpoint_design}, we design a simple checkpoint mechanism and find that it is not sufficient to match the observed checkpoint success rate data. Finally in Section~\ref{sec:true_checkpoint_design}, we propose a two-step checkpoint model design, derive necessary parameters, and verify that it matches the observed data. 

\subsection{Baseline growth} \label{sec:growth_baseline}
Since the checkpoint data in \cite{Feng2021ParallelPolymerization} measures the success rate of MT plus ends growing $2~\mu$m past the branch point at which they were nucleated, we first aim to  quantify this probability in our base model of MT dynamic instability. First, we simplify the geometry of the environment by representing the branch point at which the MT was nucleated by a point which we will refer to as a nucleation location. Then, we use our previously-published model of MT polymerization dynamics in \cite{2024NelsonMinimalMech} to quantify the percentage of MTs which grow at least $2~\mu$m from nucleation. Numerical simulations of over 1,800 MT lifetimes, from nucleation to full catastrophe, revealed that approximately $90\%$ of plus ends grow at least $2~\mu$m past their nucleation location (Appendix Figure~\ref{fig:plus_end_growth}). We use $\Delta y$ to denote the distance in microns that the MT plus end grows from the nucleation site. With this notation, the baseline growth probability for the checkpoint mechanism can be written as $\mathbb{P}(\Delta y\geq 2) \approx 0.9$. Any checkpoint mechanism applied to new MT nucleations will further reduce this probability that a MT successfully exits the nucleation site.

\subsection{Environment definition}
\label{sec:environment_definitions}

In \cite{Feng2021ParallelPolymerization}, checkpoint success rates are reported for all exit directions in both an environment with primarily MEO polarity and an environment with mixed polarity. To test the probability of MT exit success based on different nucleation mechanisms, we must define each of these environments. In the control environment, roughly $90\%$ of microtubules are MEO. On the other hand, neurons with Patronin or kinesin-2 knockdown exhibit a mixed environment, where roughly $50\%$ of microtubules are MEO. We define $\mathbb{P}(\text{PreMT}^{+/-}_\text{Ctrl/Pat})$ to be the probability that a local pre-existing microtubule is oriented PEO (+) or MEO (-) in the control (Ctrl) or Patronin (Pat) polarity environments and record these experimental parameters in Table~\ref{tab:model_parameters}. We will use these probabilities to inform the expected success rates of our checkpoint mechanism designs.

\begin{figure}
    \centering
    \includegraphics[]{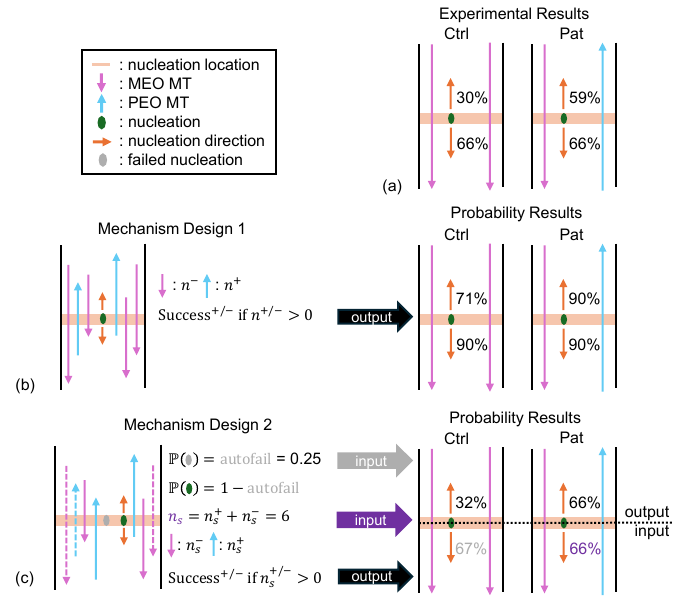}
    \caption{Checkpoint mechanism data, design and validation: (a) experimental checkpoint success rates data from \cite{Feng2021ParallelPolymerization} for PEO and MEO nucleation directions in control (Ctrl) and Patronin knockdown (Pat) background environments; (b) simple mechanism design and calculated success rates which do not match experimental data, (c) proposed mechanism design and derivation with calculated success rates which match experimental data.}
    \label{fig:checkpoint_construction}
\end{figure}

\subsection{Simple mechanism design} \label{sec:simple_checkpoint_design}

We first consider a simple design for the checkpoint mechanism, which assumes that a new nucleation will survive the checkpoint if there is at least one local, pre-existing MT with the same polarity as the new nucleation. Therefore the probability of a PEO $(+)$  or MEO $(-)$ nucleation exiting the checkpoint successfully $2~\mu$m from the nucleation location is 
$$\mathbb{P}(\text{success}^{+/-}) = \mathbb{P}(\Delta y \ge 2)\mathbb{P}(n^{+/-}\ge 1)$$
where $n^{+/-}$ is the number of local, pre-existing PEO or MEO MTs at that site. The expression $\mathbb{P}(n^{+/-}\geq 1)$ is thus equal to 1 minus the probability that all local MTs have the opposite polarity from the newly-nucleated MT. We can re-write this probability as: 
\begin{equation}\label{eq:matching_prob_simple}
\mathbb{P}(n^{+/-}\ge 1) =(1-\mathbb{P}(n^{-/+} = n)) = (1-(\mathbb{P}(\text{PreMT}^{-/+}_\text{Ctrl/Pat}))^n)
\end{equation}
where $n = n^+ + n^-$, i.e. the total number of local, pre-existing MTs.

Since studies have observed 10-20 MTs on average in a cross-section of a dendrite \cite{Kim2012,Zhang2024}, we approximate that there are $n = 15$ existing MTs at a nucleation site and test the probability that a MT survives this checkpoint mechanism. The results of this calculation in each polarity environment and for each nucleation direction given the experimental measurements in Section~\ref{sec:environment_definitions} are shown in Figure~\ref{fig:checkpoint_construction}b. Notably, these success probabilities do not match those from the experimental data illustrated in Figure~\ref{fig:checkpoint_construction}a. In particular, all estimated exit success rates are higher and the difference between MEO and PEO rates in the Ctrl environment is less pronounced than is observed in the experimental checkpoint success rates. This indicates that this simple checkpoint mechanism assumption is not sufficient to reproduce the fate of newly-nucleated MTs at dendrite nucleation sites.

\subsection{Proposed mechanism design} \label{sec:true_checkpoint_design}

We next propose a more complex checkpoint mechanism design to better reproduce the observed MT exit success rates while preserving the dependence of this mechanism on pre-existing, local MTs. As noted in Section~\ref{sec:simple_checkpoint_design}, this mechanism should generate lower overall success rates and a larger difference between MEO and PEO nucleation success in the Ctrl environment compared to the simple mechanism design.

One adjustment is that there is some chance that MT nucleations fail independently of local MT content at that site. This adjustment is biologically realistic, as we often observe very short-lived microtubules after nucleation. Mathematically, changing the model to allow for nucleation failure independently of local MT content has the potential to account for the low checkpoint success rate for MEO MTs observed in the Ctrl environment ($66\%$, see Figure~\ref{fig:checkpoint_construction}a), where almost all pre-existing MT content matches the orientation of MEO nucleated MTs. Another possible adjustment is that only a subset of the local MTs at a nucleation site can inform the checkpoint mechanism. This is plausible, since newly-nucleated MTs may only be able to observe a limited number of pre-existing MTs, due to spatial constraints. Mathematically, decreasing the number of local MTs that influence MT alignment is likely to decrease all success rates and to increase the difference between MEO and PEO nucleations in the Ctrl environment (see Equation~\eqref{eq:matching_prob_simple}).

These two adjustments suggest a checkpoint mechanism where the probability of checkpoint success for a PEO or MEO nucleated microtubule is given by:

\begin{equation}\label{eq:checkpoint_prob}
\mathbb{P}(\text{success}^{+/-}) = \mathbb{P}(\Delta y \ge 2) (1-\mathbb{P}(\text{autofail}))\mathbb{P}(n_s^{+/-}\ge 1)
\end{equation}
where $\mathbb{P}(\text{autofail})$ is an automatic probability of failing the checkpoint regardless of the orientation of pre-existing MTs, and $n_s = n_s^+ + n_s^- \le n$ is the number of MTs (selected from the local pre-existing MTs) that inform the checkpoint success. Similarly to Equation~\eqref{eq:matching_prob_simple}, the expression for observing a parallel pre-existing MT becomes
\begin{equation}
    \mathbb{P}(n_s^{+/-}\geq 1) = 1-(\mathbb{P}(\text{PreMT}^{-/+}_\text{Ctrl/Pat}))^{n_s}.
\end{equation}

To parameterize this mechanism, we need to determine parameters $\mathbb{P}(\text{autofail})$ and $n_s$ that match the biological observations.
First, we use the MEO success rate in the Ctrl environment to determine $\mathbb{P}(\text{autofail})$ (Figure~\ref{fig:checkpoint_construction}c, input in grey). Assuming that newly-nucleated MTs observe at least 2 local existing MT (i.e., $n_s\geq 2$), we approximate that $1-(\mathbb{P}(\text{PreMT}^{+}_\text{Ctrl}))^{n_s} \approx 1$ and thus $\mathbb{P}(\text{autofail})\approx 0.25$. Using this value and the observed success probability for MEO microtubules in the mixed environment (Figure~\ref{fig:checkpoint_construction}c, input in purple), we find that $n_s\approx 6$.

Finally, we validate this mechanism design and parameterization by calculating the resulting probability of success for PEO MT nucleations in both polarity environments and comparing to the biological data. The output probabilities are shown in Figure~\ref{fig:checkpoint_construction}c in black and match the observed checkpoint success rates shown in Figure~\ref{fig:checkpoint_construction}a from \cite{Feng2021ParallelPolymerization} remarkably well. We therefore implement this checkpoint mechanism design with an automatic failure probability and survival dependent on observing a pre-existing, parallel MT from a subset of local MTs in our spatial mathematical model of this system in Section~\ref{sec:math_model}. The derived model parameters are presented in Table~\ref{tab:model_parameters}.

\section{Mathematical model design}
\label{sec:math_model}

In the previous sections, we used experimental results to determine the underlying  mechanisms which dictate the polarity of new nucleations and which control the growth of these new MTs based on a checkpoint. We next seek to understand how these two mechanisms contribute to regulation of MT polarity in a linear domain. We develop a mathematical model of MT dynamics in a linear spatial domain, where the polarity regulation mechanisms are implemented at periodically-spaced nucleation locations. In Section~\ref{sec:spatial_model}, we describe our construction of a spatially-explicit model, and in Section~\ref{sec:mt_model}, we summarize our prior stochastic model of MT growth dynamics at both MT ends. In Section~\ref{sec:feedback_mech}, we present our model implementation of the nucleation feedback mechanism identified experimentally in Section~\ref{sec:nucleation_bio_data}. In Section~\ref{sec:checkpoint_mech_general}, we detail our model construction of the checkpoint mechanism as derived in Section~\ref{sec:checkpoint_mechanism}. Finally, in Section~\ref{sec:algorithm}, we describe the algorithm used to simulate this model. Table~\ref{tab:model_parameters} presents key parameters that are used in the stochastic model and experimentally-observed quantities that inform our choices.

\subsection{Spatial model design}
\label{sec:spatial_model}

To capture MT dynamics in neuronal regions, we model MT dynamics using a spatially-explicit stochastic model in a linear domain. Within this domain structure, we denote periodic sites in which all nucleation events and regulation mechanisms take place. We refer to these sites as nucleation locations. This simple, linear domain is of particular interest for modeling because it is representative of cellular domains such as neuronal dendrites and axons. In particular, we scale our domain to resemble the primary trunk of a \textit{Drosphila} dendrite as we parameterize our model using experimental results from this system. 

We orient our domain vertically with the bottom domain boundary representing the cell body and the top domain boundary representing the neurite tip (Figure \ref{fig:mt_spatial_model}a). To explore relevant behaviors in our stochastic model, we consider two different domain lengths. In the first domain, we model a single nucleation location in a domain of length $20~\mu$m (see Results Section~\ref{sec:single_nucloc}). In the second domain, we represent six equally-spaced nucleation locations in a domain of length $70~\mu$m (see Results Sections~\ref{sec:full_trunk}-\ref{sec:maintenance_statistics}).  In both domains, we maintain a distance of $10~\mu$m between each nucleation location or domain boundary (cell body or tip). The longer spatial domain is representative of a \textit{Drosophila} sensory neuron dendritic trunk, where there are typically fewer than ten nucleation locations, and the distance between consecutive nucleation sites is approximately $10~\mu$m (see example in Figure \ref{fig:biodata}a). With these two configurations, we explore polarity behavior governed by mechanisms at a single nucleation site as well as behavior emerging from interacting nucleation locations. We confine MTs to these linear domains so that growth beyond the top and bottom boundaries is not permitted. 

As in \cite{2024NelsonMinimalMech}, we assume a constant number of MTs. This is a reasonable assumption in healthy settings, where nucleation is known to be heavily controlled in the cell \cite{chen2012axon,hertzler2020kinetochore}. Therefore, in our model, when a MT completely catastrophes, a new MT nucleates at a nucleation site selected at random. We populate the shorter domain with 20 MTs and our longer domain with 60 MTs. In each simulation, the MTs have a characteristic length of $20~\mu$m, based on the model of MT growth and shrinking described in Section~\ref{sec:mt_model}. Our choices of MT number and length were based on experimental information on the number of MTs typically expected in a cross section of a dendrite. While there is little data accurately quantifying the total MT number or length \textit{in vivo} in neurons, studies have observed 10-20 overlapping MTs on average in a cross-section of a dendrite \cite{Kim2012,Zhang2024}. Our model set-up generates 9 and 13 MTs on average in cross-sections of the shorter and longer domains respectively, with most cross-sections in both cases consistent with the realistic range of 10-20 MTs. A larger number of MTs (than would be proportional to the domain length) is required in the shorter domain in order to match the target number of MTs in each cross-section. These choices of MT numbers and length result in the emergence of overlapping MTs in cross-sections of the model neurites, with the cytoskeleton often continuously spanning the vertical length of the domain, as can be seen in the Supplemental Videos.

\subsection{Stochastic model of microtubule growth dynamics}
\label{sec:mt_model}
\begin{figure}
    \centering    \includegraphics[width=\linewidth]{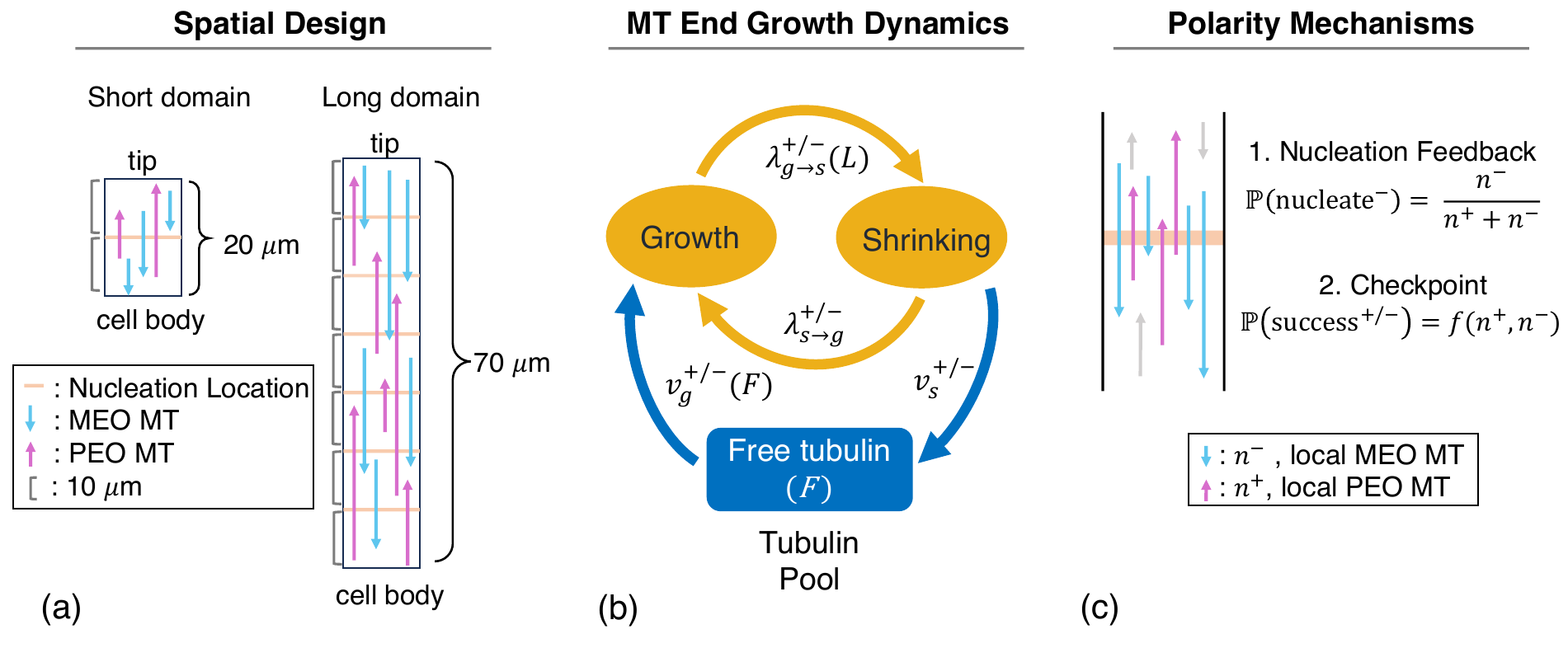}
    \caption{Schematic of the model and mechanism designs. Panel (a) illustrates our short and long domains, which are defined by the number of nucleation locations included in the spatial model. Panel (b) shows the MT end states (growth or shrinking) and how the the free tubulin pool depends on MT growth dynamics. Panel (c)	 depicts the two polarity mechanisms that depend on local MT content, $n^{+/-}$.}
    \label{fig:mt_spatial_model}
\end{figure}

Given the stochastic nature of MT growth and shrinking dynamics, our mathematical model must account for these behaviors to capture MT turnover dynamics. We use a stochastic continuous-time Markov chain (CTMC) model of MT dynamics that we developed in \cite{2024NelsonMinimalMech}, where we explored mechanisms that can regulate MT length. This model was parameterized to experimental data from \textit{Drosophila} dendrites and recapitulates MT dynamics observed \emph{in vivo}. In this model, we considered a one-dimensional domain and modeled MTs using an ordered pair of coordinates, which represent the positions of the minus and plus ends of each MT at time $t$. We assume a fixed number of MTs, $N$, throughout the simulation, and each MT end is in one of two states: growth, where the MT end is polymerizing at speed $v_g^{+/-}$, or shrinking, where the MT end is depolymerizing at speed $v_s^{+/-}$. Microtubule ends transition between these growth and shrinking states. We define $\lambda_{s\rightarrow g}^{+/-}$ to be the rate of switching from shrinking to growth (an event called rescue), while $\lambda_{g\rightarrow s}^{+/-}$ refers to the rate of switching from growth to shrinking (an event called catastrophe). The plus or minus signs in this notation refer to the speeds or switching rates at the plus and minus ends of the MTs, respectively. These MT end dynamics are illustrated in the model cartoon in Figure~\ref{fig:mt_spatial_model}b.

Our original CTMC model of MT growth dynamics explored how two length-limiting mechanisms, length-dependent catastrophe and limited tubulin availability, impact quantities such as MT lengths and distributions of MT speeds \cite{2024NelsonMinimalMech}. To model the length-dependent catastrophe mechanism, the catastrophe rates are set as a function of microtubule length, $L$, such that 
\begin{equation*}
    \lambda_{g\rightarrow s}^{+/-}(L) = \max\left( \lambda_{\text{min}},\tilde{\lambda}_{g\rightarrow s}^{+/-} + \gamma(L - L_0)\right).
\end{equation*}
In this function, $\gamma$ scales the impact of the length-dependent catastrophe mechanism based on the difference between the MT length $L$ and the characteristic length $L_0$. In the other length-limiting mechanism, tubulin (the fundamental building block of MTs) is a resource that is likely limited inside cells. Thus, in the limited tubulin availability mechanism, we assume that the MT end growth velocities $v_g^{+/-}$ are functions of the amount of free tubulin, $F$, in our system. We also assume that the total amount of tubulin, $T_{tot}$, is conserved within the system between free tubulin and tubulin in microtubules. Microtubule growth depletes the amount of free tubulin, $F$, while shrinking events replenish $F$. When MTs grow in excess of the amount of free tubulin, the growth velocities are limited by the tubulin availability. These length-regulating mechanisms are illustrated in Figure~\ref{fig:mt_spatial_model}b and further details including their mathematical model implementation are given in~\cite{2024NelsonMinimalMech}. Parameter values are specified in Table~\ref{tab:model_parameters}.

\subsection{Nucleation feedback mechanism}
\label{sec:feedback_mech}

Given the biological findings in Section~\ref{sec:nucleation_bio_data}, we investigate the nucleation feedback mechanism, which dictates the orientation of new microtubule nucleations. When a MT undergoes complete catastrophe in our model, we nucleate a new MT by randomly selecting a nucleation site. As defined previously, $n^{+/-}$ represents the number of MTs present locally at the selected nucleation location with PEO and MEO orientation, respectively (Figure~\ref{fig:mt_spatial_model}c). The fraction of local MTs which are MEO defines the probability that the new MT is nucleated MEO such that $\mathbb{P}(\text{nucleate}^{-})= \frac{n^{-}}{n^-+n^+}$. If no MTs cross the chosen nucleation location at that time point, we set $\mathbb{P}(\text{nucleate}^{-}) =  \frac{1}{2}$. If nucleation feedback is not implemented in our simulation, we also set $\mathbb{P}(\text{nucleate}^{-}) =  \frac{1}{2}$. 

We then determine the direction of nucleation by drawing from a Bernoulli distribution with probability $\mathbb{P}(\text{nucleate}^-)$. The MT minus end is then initialized at the nucleation location, while the plus end is initialized one seed length above or below the nucleation location for PEO or MEO nucleations, respectively. We chose a small, non-zero seed length in order to record the polarity decision using the MT end locations (Table~\ref{tab:model_parameters}).

\subsection{Checkpoint mechanism}
\label{sec:checkpoint_mech_general}
We implement the checkpoint mechanism based on the design validated in Section~\ref{sec:true_checkpoint_design} and we apply it to all new MT nucleations in our spatial model of MT polarity. After a MT is nucleated, we first apply an automatic failure probability by drawing from a Bernoulli distribution with probability $\mathbb{P}(\text{autofail})=0.25$. If the MT autofails, then it is set to completely catastrophe and re-nucleate in the next time step. Otherwise, we draw $n_s = 6$ MTs with replacement from the set of MTs that cross the relevant nucleation site. If any of the selected MTs have the same orientation as the newly-nucleated MT, the new MT survives. If none of the selected MTs have the same orientation, the new nucleation fails the checkpoint and is set to completely catastrophe and re-nucleate in the next time step. If there are no MTs at the selected nucleation location, the newly-nucleated MT behaves as if it observes a pre-existing MT with the same orientation, for simplicity.

We validate this checkpoint mechanism implementation by tracking MT checkpoint success rates in simulations of our spatial model. We simulate our model with the long domain configuration initialized with 90\% MEO MTs (the Ctrl environment) and with 50\% MEO MTs (the Pat environment). We record the average checkpoint success rate at each nucleation location for each direction. Supplemental Figure~\ref{fig:checkpoint_model_success} illustrates the average model success rates, compared with the data from \cite{Feng2021ParallelPolymerization}. Our results show strong agreement of the model and experimental success rates, further validating our proposed checkpoint mechanism design in Section~\ref{sec:true_checkpoint_design}.The spatial nucleation mechanisms that influence microtubule polarity are illustrated in Figure \ref{fig:mt_spatial_model}c.

\subsection{Algorithm for model implementation}
\label{sec:algorithm}
To simulate the spatial model, we first initialize the fixed number of MTs in the domain. At first, all MTs have the same length, are uniformly distributed through the domain, and the proportion starting MEO is set to the desired starting polarity. Typically, we set this proportion to be $50\%$. We use a time step of $1$ second. Then, for every second between the starting and ending times, we iterate through a series of modeling steps as follows:
\begin{enumerate}
    \item \textbf{Polymerization Step}: Update MT end locations and length based on their polymerization and depolymerization dynamics as outlined in \cite{2024NelsonMinimalMech}. Reposition any MT ends intended to grow past a domain boundary and set them to the domain boundary. Count the number of MTs that have undergone full catastrophe (i.e. shrunk to length $\leq 0$) to inform the MT nucleation step.
    \item \textbf{Checkpoint Step}: If the checkpoint mechanism is not turned on (or if this is the first time step), skip this step. Otherwise, for each MT nucleated in the previous time step, perform the checkpoint. First, draw a uniform random number on $[0,1]$ and compare it to the automatic failure probability. If the MT fails, remove the MT and count it for the nucleation step. If the MT passes the automatic failure, select $n_s$ MTs with replacement from those that cross this MT's nucleation location. If no selected MTs match this MT's nucleation direction, remove the MT and count it for the nucleation step. If at least one selected MT matches or there are no local MTs, do nothing.
    \item \textbf{Nucleation Step}: Add the numbers of MTs which were removed during the polymerization and checkpoint steps in the current time step. For each MT removed, nucleate a new MT. To nucleate a new MT, first pick a nucleation location uniformly at random. Then, determine the probability of the new MT being oriented MEO. If nucleation feedback is not implemented or there are no MTs crossing the chosen nucleation location, the probability of nucleating in the MEO direction  is 50\%. Otherwise, the probability of nucleating MEO is equal to the proportion of pre-existing MTs crossing the chosen nucleation location which have MEO polarity. Draw a uniform random number on $[0,1]$ and compare to the MEO nucleation probability to determine if the new MT is nucleated in the MEO or PEO direction. The new MT has its minus end set at the nucleation location, while its plus end is set one seed length above or below the nucleation location for PEO and MEO nucleations, respectively. 
\end{enumerate}

To investigate how polarity is established and maintained in these different domains, we run all simulations for 5 days. This timescale corresponds to the average lifespan of the \textit{Drosophila} larva, the organism which informs the parameterization of our model \cite{Kargbo-Hill2012}. Since biased MT arrays are maintained in neurons over the lifetime of these organisms, we are interested in exploring the emergence, maintenance, and stability of biased polarity distribution on this timescale. We are particularly interested in what we will refer to as ``fully-biased polarity" distributions, where all simulated MTs exhibit the same polarity, i.e. either 100\% or 0\% MEO. When nucleation feedback is implemented, departure from a fully-biased polarity distribution is only possible if a new MT is nucleated with a random orientation at a nucleation site where no other MTs cross through at that time point. When checkpoint is implemented, temporary departure from a fully-biased polarity is allowed for a single time step since the checkpoint mechanism is applied to nucleations after one growth step. As is the case for the nucleation feedback mechanism, departure from a fully-biased polarity only persists when the MT was nucleated in a random direction at a nucleation site with no other MTs crossing through at that time point. We discuss results derived from simulations with this model algorithm in Section~\ref{sec:results}.

\begin{table}[!h]
    \centering
 \begin{footnotesize}
    \caption{Parameters used in the spatial stochastic model from Section~\ref{sec:math_model} and experimentally-observed quantities that inform our modeling choices. Parameters marked with $\dagger$ are preliminary estimates from the Rolls lab and parameters marked with $\ddagger$ are found through the parameterization procedure outlined in \cite{2024NelsonMinimalMech}.}
    \begin{tabular}{r|c|c|c} 
           \hline \hline 
         \textbf{MT growth model parameters \cite{2024NelsonMinimalMech}} & \textbf{Notation} &\multicolumn{2}{c}{\textbf{Value}} \\ \hline
        Characteristic MT Length & $L_0$ & \multicolumn{2}{c}{$20\mu$m}\\
                 Number of MTs & $N$ & \multicolumn{2}{c}{20 or 60 } \\
        Total available tubulin & $T_{tot}$ & \multicolumn{2}{c}{$\frac{L_0}{35}\times50N \mu$m}\\
        Length-dependence of catastrophe rate & $\gamma$ & \multicolumn{2}{c}{$0.01~(\mu \text{m min})^{-1}$ } \\
        Minimum catastrophe rate & $\lambda_{\text{min}}$ & \multicolumn{2}{c}{$0.05/\text{min}$} \\
        \hline
        \textbf{Experimentally-informed quantities}& \textbf{Notation}& \textbf{Plus end}& \textbf{Minus end}\\
        \hline
       Average polymerization speeds \cite{Feng2019} & $v_g^{+/-}$ &  $6~\mu$m/min & $0.75~\mu$m/min\\
        Depolymerization speeds $\dagger$ & $v_s^{+/-}$ &  $6~\mu$m/min & $3.5~\mu$m/min\\
        Average growth-phase duration\cite{Feng2019} & $\overline{\tau}_g^{+/-}$ & 2 min & 4 min \\
        Catastrophe rates at length $L_0$ \cite{Feng2019} & $\tilde{\lambda}_{g\rightarrow s}^{+/-} = 1/\overline{\tau}_g^{+/-}$ &  $0.5$/min & $0.25 $/min\\
        Rescue rates $\ddagger$ & $\lambda_{s\rightarrow g}^{+/-}$  &  $0.2811$/min & $0.67 $/min\\
    \hline\hline
 \textbf{Spatial model parameters} &  \multicolumn{3}{c}{\textbf{Value}} \\\hline
         Number of nucleation locations &  \multicolumn{3}{c}{1 or 6 } \\
         MT initial length &  \multicolumn{3}{c}{$15\mu$m }\\
         MT seed length &  \multicolumn{3}{c}{$0.1 \mu$m }\\
         Length between nucleation locations & \multicolumn{3}{c}{$10\mu$m} \\
          Domain length & \multicolumn{3}{c}{$20\mu$m or $70\mu$m} \\
          \hline \hline
        \textbf{Checkpoint environment definitions} & \textbf{Notation} &\textbf{Plus end} & \textbf{Minus end}  \\ \hline
        Control environment & $\mathbb{P}(\text{PreMT}^{+/-}_{\text{Ctrl}})$& 10\% & 90\% \\
        Patronin knockdown environment & $\mathbb{P}(\text{PreMT}^{+/-}_{\text{Pat}})$& 50\% & 50\%  
        \\ \hline
         \textbf{Experimental checkpoint success rates \cite{Feng2021ParallelPolymerization}} & \textbf{Notation} &\textbf{Plus end} & \textbf{Minus end}  \\ \hline
       Control environment& Ctrl  &30\% &66\%\\
        Patronin knockdown environment& Pat& 59\% & 66\% \\
        \hline\hline
        \textbf{Nucleation mechanism parameters} & \textbf{Notation}& \multicolumn{2}{c}{\textbf{Value} }\\\hline
         Number of +/- MTs at nucleation location & $n^{+/-}$ & \multicolumn{2}{c}{Varies} \\
         MEO nucleation probability & $\mathbb{P}(\text{nucleate}^-)$ & \multicolumn{2}{c}{$\frac{n^{-}}{n^++n^-}$}\\
        Growth $\geq 2\mu$m probability &$\mathbb{P}(\Delta y \geq 2)$ & \multicolumn{2}{c}{0.9 } \\
        Auto fail probability &$\mathbb{P}(\text{autofail})$ & \multicolumn{2}{c}{0.25 } \\
        Number of selected MTs for checkpoint & $ n_s$ & \multicolumn{2}{c}{6 }\\
        \hline \hline
    \end{tabular}
    \label{tab:model_parameters}
    \end{footnotesize}
\end{table}

\section{Results}
\label{sec:results}
In Sections~\ref{sec:checkpoint_mechanism} and~\ref{sec:math_model}, we outlined a mathematical framework that incorporates two \textit{in vivo} mechanisms thought to be important for the generation of biased MT polarity in living neurons: nucleation feedback and a post-nucleation checkpoint mechanism. While these nucleation mechanisms have been studied experimentally, it is unknown precisely how nucleation impacts polarity dynamics and which mechanisms are sufficient to establish and maintain biased polarity. Here, we combine these two nucleation mechanisms together with stochastic MT growth dynamics in a spatial model that investigates the establishment and maintenance of MT polarity in different domain configurations. This mathematical modeling framework allows us to explore these mechanisms in a systematic way by simulating the four possible scenarios with each mechanism turned ``off" or ``on", i.e. neither mechanism, checkpoint-only, feedback-only, or both mechanisms. As noted in Section~\ref{sec:feedback_mech}, when simulating with the nucleation feedback mechanism turned off (i.e. for neither or checkpoint-only combinations) we employ a fixed, 50\% probability of new MTs nucleating with either polarity.

Since MT nucleation occurs in a variety of cell types and cell lengths, exploring polarity behavior in both a short and long domain design can give useful information about polarity maintenance. In Section~\ref{sec:single_nucloc}, we study how nucleation mechanisms determine polarity bias with a single nucleation location in a short domain over a 5-hour time span. In Sections~\ref{sec:full_trunk}-\ref{sec:maintenance_statistics}, we study the interactions between MTs across multiple nucleation locations to establish and maintain polarity bias across a longer domain over 5-day simulations. For each domain type, we initialize our mathematical model with MTs that have mixed polarity, i.e. 50\% MEO, to investigate the mechanisms that influence polarity establishment. We are particularly interested in the emergence of fully-biased polarity distributions, where all simulated MTs exhibit the same polarity, i.e. either 100\% or 0\% MEO.

\subsection{Nucleation feedback and checkpoint mechanisms establish local bistable polarity on different timescales}\label{sec:single_nucloc}

\begin{figure}
    \centering
    \includegraphics[width=\linewidth]{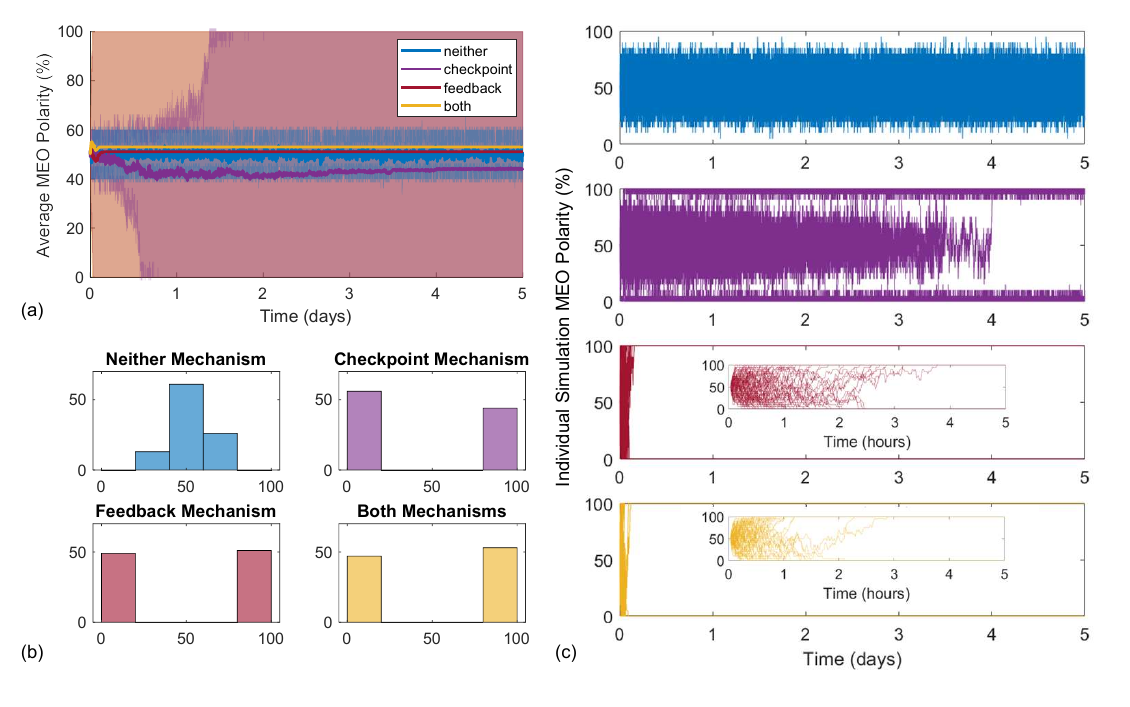}
    \caption{
    Microtubule polarity results for 100 runs of 5-day simulations with a single nucleation location and $N = 20$ MTs in a $20~\mu$m domain. The simulations are initialized with 50\% MEO polarity for each combination of mechanisms: neither (blue), checkpoint-only (purple), feedback-only (red), and both (yellow). The average MEO polarity over time is shown in panel (a) as a solid curve, with a cloud representing the interquartile range. Panel (b) shows the polarity distribution at the end of 5 days, and (c) shows the MEO polarity percentage for each simulation over time. The inset figures for feedback-only (red) and both mechanisms (yellow) show how the MEO polarity percentage changes on a shorter time scale of 5 hours.
    }
    \label{fig:singlenucloc}
\end{figure}

We first study how biased polarity can be established by the two MT nucleation mechanisms in a short domain with a single nucleation location as outlined in Section~\ref{sec:spatial_model}. For each of the four mechanism combinations, we simulate 100 realizations of the spatial model for 5 days each to understand how each nucleation mechanism contributes to polarity establishment. 

To study whether biased MT polarity emerges, we track the proportion of MTs which are MEO over time. The results in Figure~\ref{fig:singlenucloc} show how average MEO polarity changes over time for neither mechanism (blue), checkpoint-only (purple), feedback-only (red), and both mechanisms (yellow). The solid curves in Figure~\ref{fig:singlenucloc}a illustrate that the average MEO polarity for each mechanism combination remains close to 50\%, while the interquartile range (IQR) clouds reveal different behaviors. Polarity has less variance in the neither mechanism case, while the IQR is wide for the checkpoint-only, feedback-only, and both mechanisms. To better understand this behavior, we illustrate the evolution of MT polarity through time for each individual simulation in Figure~\ref{fig:singlenucloc}c. The top panel shows that the scenarios with neither mechanism implemented largely yield 40\%-60\% MEO polarity. With the checkpoint-only mechanism, only a few model runs reach a fully-biased polarity state of 100\% or 0\% MEO polarity within one day. After approximately four days, all runs reach a roughly fully-biased polarity, with some variability close to the fully-biased polarity states due to the algorithm implementation.  In contrast, all simulations of the feedback-only and both mechanisms cases in the bottom two panels of Figure~\ref{fig:singlenucloc}c reach a fully-biased polarity state within 5 hours (see figure insets). This shorter timescale is also relevant biologically, since experiments have shown that microtubule polarity is established within the first 24 hours of the \emph{Drosophila} larval stage \cite{Kargbo-Hill2012}.

In Figure~\ref{fig:singlenucloc}b, we illustrate the MEO polarity distributions at the end of the simulations, which show the range of polarity behaviors in each mechanistic scenario. Neither mechanism results in a unimodal normal-like distribution centered around 50\% MEO polarity. Again, this shows that we do not achieve a fully-biased polarity state with neither mechanism. However, for feedback-only, checkpoint-only, and both mechanisms, the final MEO polarity distributions are nearly identical and clearly bimodal, with all simulations achieving either 0\% MEO polarity or 100\% MEO polarity by the end of each simulation. We show sample simulations with both nucleation mechanisms implemented which result in a fully-biased MEO and PEO polarity state in Supplemental Video 1. While the feedback-only and checkpoint-only mechanisms can result in fully-biased polarity, the time at which MTs reach robust complete polarization differs greatly between these mechanisms (hours versus days). The time course plots in Figure \ref{fig:singlenucloc}c illustrate that feedback alone and both mechanisms are very effective at achieving biased polarity for a single nucleation location on a short time-scale, and suggest that the system exhibits bistability on this spatial scale.

\subsection{Nucleation feedback mechanism establishes global bistable polarity}
\label{sec:full_trunk}

\begin{figure}
    \centering
    \includegraphics[width=\linewidth]{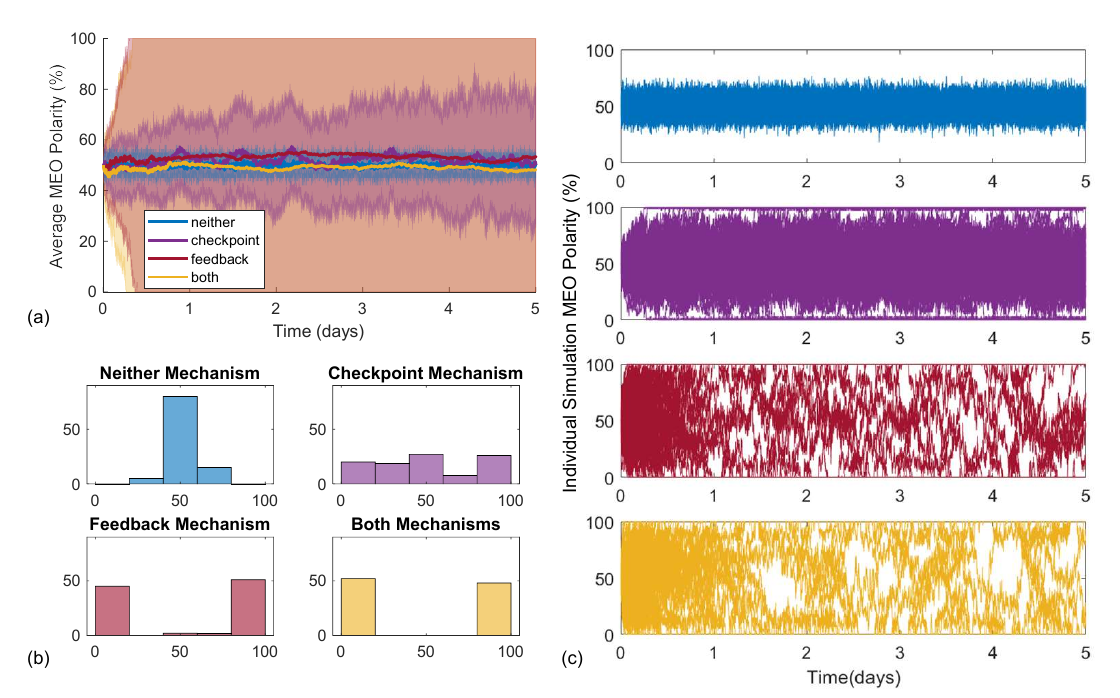}
    \caption{Microtubule polarity results for 100 runs of 5-day simulations with six nucleation locations and $N = 60$ MTs in a $70~\mu$m domain. The simulations are initialized with 50\% MEO polarity for each combination of mechanisms: neither (blue), checkpoint-only (purple), feedback-only (red), and both (yellow). The average MEO polarity over time is shown in panel (a) as a solid curve, with a cloud representing the interquartile range. Panel (b) shows the polarity distribution at the end of 5 days, and (c) shows the MEO polarity percentage over time for each simulation.}
    \label{fig:full_trunk}
\end{figure}

Biased polarity is especially critical in longer cells, where directional transport is key for the long-range movement of protein cargo. To understand polarity establishment and maintenance in such environments, we investigate the spatial stochastic model in a longer domain with 6 nucleation locations and 60 MTs initialized with mixed polarity, as outlined in Section~\ref{sec:spatial_model}. Here, we seek to understand how MTs across multiple nucleation sites interact and how biased polarity is maintained and established. As in Section~\ref{sec:single_nucloc}, we track the proportion of MTs which are MEO over time in each combination of mechanisms. 

The results of 100 simulations of each model scenario are shown in Figure~\ref{fig:full_trunk}. As in Figure~\ref{fig:singlenucloc}a, Figure~\ref{fig:full_trunk}a shows that the average polarity remains mixed at $50\%$ MEO, while the IQR increases throughout time for each case except for the neither mechanism case. Figure~\ref{fig:full_trunk}c shows the individual polarity time course for each mechanistic scenario. Compared to the single nucleation case in Figure~\ref{fig:singlenucloc}c, the individual polarity trajectories are harder to interpret and do not exhibit clear bistability. Throughout time, we observe a range of different polarity profiles, including some that achieve and stay at fully-biased polarity states, and some with mixed polarity. In particular, we observe tight regulation of mixed polarity for neither mechanism (blue) in the top panel of Figure~\ref{fig:full_trunk}c. We find a wider polarity range for the checkpoint-only mechanism (purple), with some trajectories reaching 0\% and 100\%. In the bottom two panels of Figure~\ref{fig:full_trunk}c, only a few simulations in the feedback-only (red) and both mechanisms (yellow) cases show nonbiased polarity behavior by the end of 5 days.

We also study the distribution of polarity behavior in this longer domain at the end of the 5-day simulations in Figure~\ref{fig:full_trunk}b. Similar to the single nucleation case, we find that the neither mechanism case has a tight, unimodal distribution around 50\% MEO polarity. However, the checkpoint-only mechanism case in Figure~\ref{fig:full_trunk}b now shows an approximately uniform distribution compared to the bimodal distribution observed in Figure~\ref{fig:singlenucloc}b. Extending simulations of the checkpoint-only case to 15 days did not alter this distribution (results not shown). This indicates that, in the longer domain, the checkpoint mechanism is insufficient to generate biased polarity on these timescales. The feedback-only and both mechanisms simulations are again very similar to the shorter domain and result in primarily bimodal distributions, but now there are a few simulations resulting in non-biased polarity in each distribution.

Notably, while most of the results here, excluding the checkpoint mechanism, are similar to those obtained with one nucleation site, the timescale to achieve fully-biased polarity and the polarity trajectories through time differ significantly. Figure~\ref{fig:full_trunk}c shows that many simulations of the feedback-only and both mechanisms scenarios result in biased polarity by 24 hours. This is a much longer time scale compared to Figure~\ref{fig:singlenucloc}c, where the mechanisms that include nucleation feedback result in biased polarity in 2-3 hours. These results suggest that the domain length and the number of nucleation locations influence the timescale on which the model achieves biased polarity, but we expect that this emergent polarized MT organization is consistent even when varying nucleation site numbers.

\subsection{Global bistable polarity establishment follows spatial phase separation}
\label{sec:nuclocs_in_fulltrunk}

\begin{figure}[h]
    \centering
    \includegraphics[width=\linewidth]{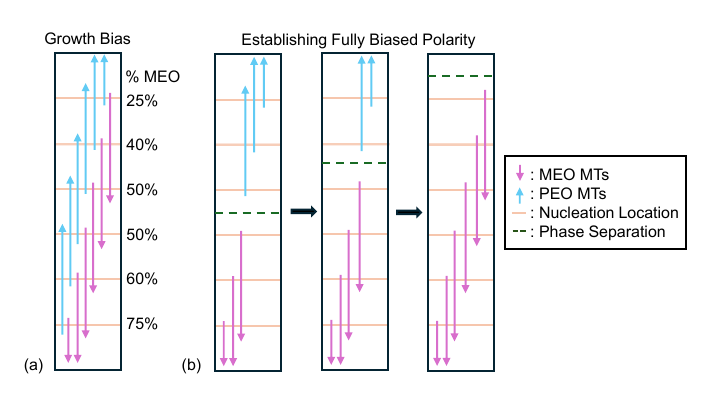}
    \caption{Schematic illustrating how polarity varies spatially across nucleation locations. Without any additional nucleation mechanism, (a) shows how MEO polarity varies spatially due to MT growth dynamics alone, with MEO polarity percentage shown for each nucleation location to the right of the figure. With nucleation mechanisms, (b) illustrates the phase separation behavior between MEO and PEO nucleation locations, where polarity is spatially-biased and consistent with the growth bias from (a). Over time, this phase separation front moves until one polarity bias is established globally. }
    \label{fig:spatial_schematic}
\end{figure}

So far, we find that the proposed nucleation mechanisms have the ability to establish biased MT polarity throughout both short and long linear cellular domains. A question remains: how do MT arrays with mixed-polarity evolve through time and space to enforce fully-biased MT polarity in a model neurite?
We thus investigate how MT nucleation mechanisms establish biased MT polarity through space in the longer domain. We begin by studying the polarity behavior at each nucleation site, where the model mechanisms influence nucleation. For each of the six nucleation locations and every time step, we focus on the MTs crossing that location and measure the MEO polarity as the proportion of the crossing MTs which are MEO. The schematic in Figure~\ref{fig:spatial_schematic}a illustrates this calculation of MEO percentages for a simple example of  MTs in the long domain.

To understand how spatial polarity bias can change over time, Figure~\ref{fig:spatial_schematic} shows representative polarity behavior similar to what is seen in simulations. Figure~\ref{fig:spatial_schematic}a illustrates an initial spatial polarity bias caused by MT growth dynamics, which we observe in all scenarios, including those with no nucleation mechanism enforced. The average polymerization speeds and growth durations in Table~\ref{tab:model_parameters} imply that plus ends have an average run length of 12 microns, compared to 3 microns  for minus ends. Therefore, plus ends are more likely to cross multiple nucleation sites and reach the ends of the neurite. For PEO MTs, this means that plus ends will travel towards the tip of the domain, while for MEO MTs, plus ends will travel towards the cell body.
This ultimately impacts the MT polarity at the nucleation sites (see Figure~\ref{fig:spatial_schematic}a) and leads to a spatial bias, with more PEO-biased nucleation locations near the tip and more MEO-biased nucleation locations near the cell body.

When nucleation mechanisms are implemented, Figure~\ref{fig:spatial_schematic}b shows how fully-biased polarity is established by first generating biased polarity locally. Our results indicate that, in a short period of time, individual nucleation sites achieve spatially-dependent biased polarity: typically, nucleation sites near the cell body are MEO and those near the tip are PEO, consistent with the growth bias in Figure~\ref{fig:spatial_schematic}a.
These biased nucleation locations quickly organize around a spatial boundary, where all nucleation locations with MEO polarity are below this boundary and all with PEO polarity are above. We refer to this spatial boundary as a ``phase separation". In all models simulations which establish fully-biased MT polarity, this phase separation moves towards either end of the domain, eventually establishing the same polarity bias across all nucleation sites. A sample layout and movement of this phase separation is depicted in Figure~\ref{fig:spatial_schematic}b. When the phase separation moves to the tip of the domain, the simulation will exhibit fully-biased MEO polarity (Figure~\ref{fig:spatial_schematic}b), while when the phase separation moves to the cell body, the simulation will exhibit fully-biased PEO polarity. Supplementary Video 2 show sample simulations of phase separation and how overall spatial polarity bias changes over time. These videos demonstrate the biased polarity at each nucleation location on a short time-scale, the development of a phase separation between the MEO and PEO MTs, and the movement of this phase separation on different timescales to achieve complete biased polarity throughout the model neurite.

\begin{figure}[h]
    \centering
    \includegraphics[width=\linewidth]{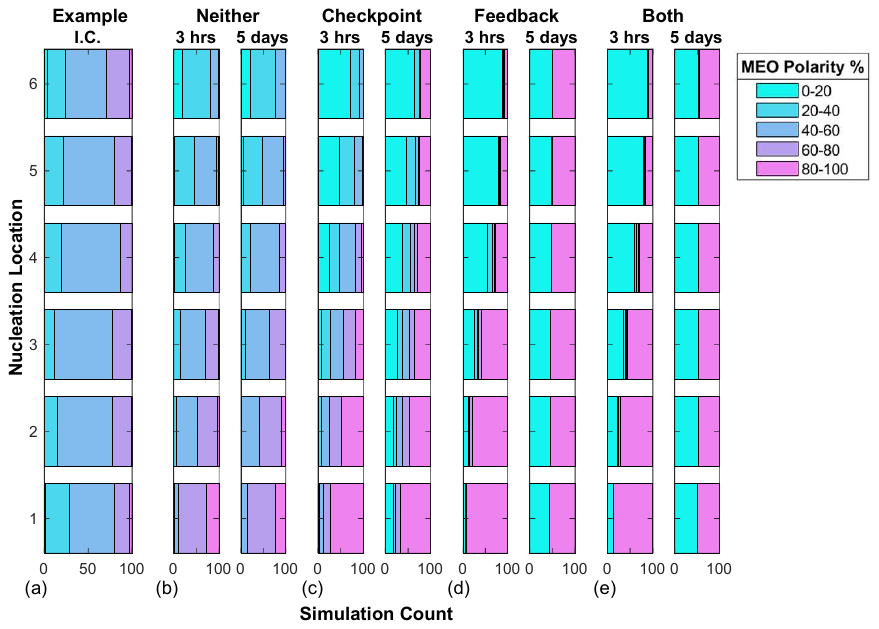}
    \caption{Distributions of MEO polarity percentages at each nucleation location. Panel (a) shows an example initial condition (I.C.) from one set of simulations. Each remaining subpanel captures the resulting polarity for one mechanism combination: (b) neither mechanism, (c) checkpoint-only, (d) feedback-only, (e) both mechanisms. For each scenario, we show the distributions of MEO polarity at 3 hours and 5 days of simulation time, with the color bar corresponding to the percentage of MEO polarity. }
    \label{fig:spatial_polarity}
\end{figure}

To further understand how polarity varies spatially and is established through time at each nucleation site, we show the distributions of MEO polarity percentages for each mechanism combination at 3 hours and 5 days of simulation in Figure~\ref{fig:spatial_polarity}b-e. We illustrate the distributions using a bar with a gradient of colors for each polarity category from blue for PEO to pink for MEO. To appreciate the change in polarity distributions over time, we illustrate the initial distribution from one set of simulations as an example initial condition in Figure~\ref{fig:spatial_polarity}a. Across all mechanisms, we observe a spatial bias in polarity at 3 hours, where most nucleation locations near the cell body exhibit a bias towards MEO polarity and most locations near the tip exhibit PEO polarity. This is evidence of the initial spatial polarity bias due to intrinsic MT growth dynamics, as in Figure~\ref{fig:spatial_schematic}a. This initial spatial bias is more pronounced when considering either or both nucleation mechanisms: in Figures~\ref{fig:spatial_polarity}c,d,e, almost all MTs at the cell body exhibit MEO bias and almost all the MTs at the tip exhibit a PEO bias. For feedback-only and both mechanisms, almost all nucleation locations are in the 0-20 or 80-100 MT polarity categories by 3 hours, which confirms that individual nucleation locations achieve biased polarity on a short timescale in these scenarios, as observed in Section~\ref{sec:single_nucloc}. The biased polarity that emerges quickly at each nucleation site is therefore robust, even in the presence of other interacting nucleation locations.

Figures~\ref{fig:spatial_polarity}b,c show that there are no significant changes after 5 days of simulation. The only apparent change between 3 hours and 5 days for these mechanism combinations are that a few simulations reach a fully-biased polarity with the checkpoint mechanism, since some nucleation sites at the tip exhibit MEO polarity and some nucleation sites at the cell body exhibit PEO polarity. For the feedback-only and both mechanisms shown in Figure~\ref{fig:spatial_polarity}d and e, respectively, the spatial bias has mostly disappeared after 5 days, with each nucleation location having fully-biased PEO or MEO polarity, and a roughly even split between these two states for the 100 model runs. This phenomenon is due to most runs becoming fully-biased to either PEO or MEO polarity, as seen in Figure~\ref{fig:full_trunk}b. Again, this complete MT polarity alignment occurs through a gradual movement of the phase separation between MEO and PEO nucleation locations to the cell body or tip. Competition at this interface often takes time, which is why bistability is achieved on a much slower timescale when starting with mixed polarity in a full neurite rather than for a single nucleation location. 

Finally, we observe that the spatial polarity behaviors emerging from the feedback-only and the both mechanisms simulations in Figs.~\ref{fig:spatial_polarity}d,e are very similar, which is consistent with the overall polarity results in Section~\ref{sec:full_trunk}. The nucleation feedback mechanism thus emerges as the key driver for achieving bistable, fully-biased polarity states in the neurite model.

\subsection{Nucleation feedback mechanism maintains global bistable polarity on long timescale}
\label{sec:maintenance_statistics}

\begin{figure}[ht!]
    \centering
    \includegraphics[width=0.8\linewidth]{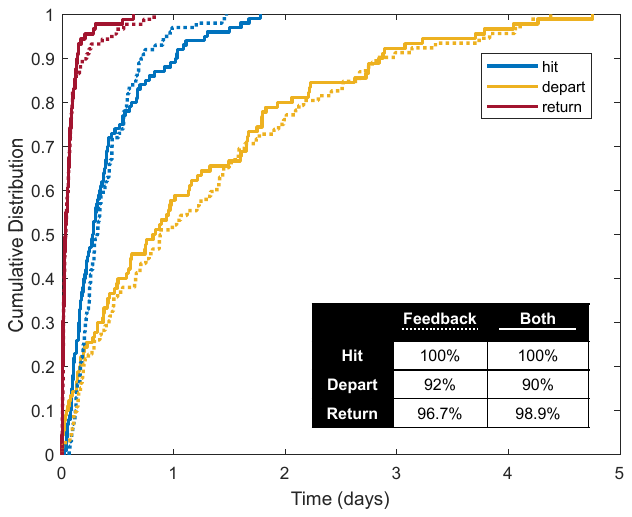}
    \caption{Cumulative distribution functions (CDFs) of data from 100 model runs for the first time to hit (blue), to depart (yellow) and to return to (red) a fully-biased MT polarity distribution (100\% or 0\% MEO) for feedback-only (dotted) and both mechanism (solid) combinations. The CDFs include data from simulations which achieved each metric, and percentages of such runs are shown for each category in the inset table.}
    \label{fig:maintenance_statistics}
\end{figure}

Biased MT polarity distributions 
are maintained for the lifetime of neurons, enabling continued and lifelong transport of intracellular proteins. Our results so far demonstrate that scenarios that include the nucleation feedback mechanism establish fully-biased polarity distributions (100\% or 0\% MEO polarity). Here, we investigate the maintenance of such biased MT polarity distributions in the feedback-only and both mechanism combinations in the long domain.

We propose measures that quantify how our model simulations reach a fully-biased MT polarity distribution. For the 100 model runs in the scenarios of interest, we track how long it takes each simulation to hit a fully-biased polarity distribution, to depart this fully-biased polarity distribution, and to return to the same distribution after leaving it. We illustrate the cumulative distribution functions for the times to hit, return, and depart from a fully-biased MT polarity distribution in Figure~\ref{fig:maintenance_statistics}. We only include data for those simulations which achieved each of these events. The inset table in Figure~\ref{fig:maintenance_statistics} shows the percentage of simulations where the event occurred in each mechanism combination.

We again observe no significant differences between simulations with feedback-only and those with both mechanisms. Thus, the nucleation feedback mechanism plays the primary role in achieving and maintaining biased MT polarity distributions in our model, and the checkpoint mechanism does not significantly impact the timing of these measures. Further, we find that most simulations achieve a fully-biased MT polarity distribution within 1 day of simulation, while all runs achieve full bias within 2 days, consistent with results from prior sections.

We observe a much longer timescale for departing a fully-biased MT polarity distribution, with some simulations taking close to 5 days to depart from such a state. In our model, departing a fully-biased MT polarity distribution is only possible when nucleation feedback is implemented and if a newly-nucleated MT is at a nucleation site with no other MTs crossing through at that time point. Examples of such a departure can be observed in panels 2 and 3 of Supplementary Video 2. Given the robust spatial coverage of MT content, this is a rare event in our simulations, which points to the biological significance of similar robust MT coverage \textit{in vivo}. Therefore, fully-biased MT polarity distributions are relatively stable in our system. This long maintenance timescale also corresponds to the average lifespan of the model organism \textit{Drosophila} larva. 

Finally, the timescale for simulations to return to the same polarity bias after departing from it is remarkably short, with most returning in just a few hours. These biased polarity distributions are thus relatively stable to perturbations in our model. Beyond the data represented in the figure, we also note that after reaching a fully-biased MT polarity, only $22\%$ of simulations of the feedback-only mechanism ever reach mixed $50\%$ polarity again, and $16\%$ do so for both mechanisms. Cumulatively, these statistics demonstrate that these mechanisms are sufficient to achieve fully-biased polarity, and that these biased polarity states are relatively stable in our stochastic model system.

\section{Discussion}
\label{sec:discussion}

The establishment and maintenance of polarized microtubule arrays in living cells is essential for long-range and long-lived intracellular transport. These polarized cytoskeletal arrays must persist even as microtubules turn over through dynamic instability and as new microtubules nucleate. Moreover, different polarizations of the microtubule arrays are observed in different regions of a neuron and these arrays have been observed to adapt in response to injury. In many neurons, including \emph{Drosophila} neurons, nucleation mechanisms have been identified that regulate microtubule polarity as nucleations occur in specific spatial locations throughout the neurite. However, how these nucleation mechanisms function and how they affect overall microtubule polarity is not well understood. Here, we used mathematical modeling and experimental validation to study adaptive nucleation mechanisms that impact polarity establishment and maintenance in neurons. 

Previous experimental results indicate that, in \textit{Drosophila} dendrites, the survival of new microtubules as they leave their nucleation region depends on the orientation of pre-existing microtubules through a checkpoint mechanism~\cite{Feng2021ParallelPolymerization}. By calculating the probabilities of success of newly-nucleated microtubules and using experimental validation, we find that the checkpoint mechanism must incorporate an automatic failure process (i.e, some nucleated microtubules are automatically rejected), and that the probability of success of new microtubules may depend on a subset of local microtubules. In addition, we carry out novel experiments that demonstrate that the observed bias for microtubule nucleations is driven by feedback from local microtubule content at nucleation locations. Experiments have suggested that this nucleation feedback could function through orientation of the nucleation sites on membranes \cite{Yalgin2015}. We thus investigate the relationship between microtubule nucleation and polarity by studying these two adaptive nucleation mechanisms which have been identified in \textit{Drosophila} neurites. 

To explore these mechanisms, we developed a spatially-explicit mathematical model of microtubule dynamics that incorporates the nucleation mechanisms, as well as the stochastic growth and shrinking at both microtubule ends \cite{2024NelsonMinimalMech}. We investigate how polarity is established and maintained spatially in a simplified linear domain, which allows us to focus on local interactions between microtubule nucleation sites. We evaluate our model on both a short domain (with one nucleation location) and a long domain (with six nucleation locations) and find that the nucleation feedback mechanism is sufficient to establish biased polarity in both cases, although the long domain consistently takes a longer time to establish this bias. We also find that the checkpoint mechanism can establish biased polarity in the short domain on a longer timescale, but it does not consistently achieve fully-biased polarity in the long, biologically-motivated domain on any timescale of relevance to our biological system.  In addition, we find that nucleation feedback alone is sufficient to establish either MEO or PEO microtubule polarity across the entire long domain, which is an emergent behavior that was not explicitly implemented in the model. This shows that this local nucleation mechanism is flexible and, since the domain is symmetric, the model can generate either type of biased polarity. In specific biological scenarios (such as axons or dendrites), we expect that other factors, such as the entrance of MTs with certain orientations from the boundaries of the neurites, will influence which polarity is chosen. Since our spatially-explicit model can capture the establishment and maintenance of either polarity orientation, our framework can be used to test experimental hypotheses. Determining the exact mechanisms that influence the establishment or reversal of a certain polarity is the subject of future work.

Even with no nucleation mechanisms in the model, we observe a spatial bias in polarity caused by the intrinsic microtubule growth dynamics, where there are more MEO microtubules near the cell body and more PEO microtubules near the tip. This spatial heterogeneity in microtubule polarity is consistent with biological observations in \textit{Drosophila} dendrites \cite{Stone2008, ori2012golgi}. Our model reveals that this spatial heterogeneity is due to the different microtubule end dynamics, which were parameterized using experimental data in \cite{2024NelsonMinimalMech}. When including nucleation mechanisms in the long neurite domain, we find that this initial spatial bias further develops into a phase separation between MEO and PEO microtubules. This is characterized by a spatial boundary between nucleation sites that have MEO (below) and sites that have PEO polarity (above). The phase separation moves as bordering nucleation locations switch between PEO and MEO polarity in a manner similar to a random walk process. Additional studies could provide further insight into this emergent behavior. This phenomenon precedes the establishment of fully-biased polarity, where the phase separation has moved to either domain boundary (cell body or tip). Once our model neurite reaches a fully-biased polarity state, we find that this polarity bias is robust and that it is maintained on timescales relevant to the lifetime of neurons.  

One limitation of our model is that we use a simple, linear domain for our simulations. To understand polarity establishment in a more complex neurite structure such as the branched dendrite of the \emph{Drosophila} neurons, we plan to extend our spatial model to study how polarity is established and maintained in a branched domain. In such a domain, additional mechanisms such as steering of microtubules by molecular motor proteins is also likely to contribute to microtubule polarity~\cite{chen2014eb1,doodhi2014mechanical}. Our model structure is also likely to be sensitive to parameters such as the microtubule length and the distance between nucleation sites, since gaps in filament content at a given nucleation site can result in no feedback from the nucleation mechanisms. 

Our framework also assumes that the number of microtubules in the domain remains fixed throughout the simulation. However, experimental data indicates that nucleation is up-regulated after injury  \cite{Stone2010,chen2012axon,Nguyen2014gamma_polarity,Weiner2020} and it is likely that additional shorter microtubules can provide more feedback at nucleation sites. In addition, microtubules may also be nucleated at domain boundaries or enter the domain from the boundary. In \textit{Drosophila} dendrites, complete axon removal leads to the conversion of a dendrite into an axon, which requires that all microtubules undergo polarity reversal \cite{Stone2010}. In future work, it would be interesting to use this mathematical modeling framework to determine what types of inputs and feedback can result in MT polarity reversals after axon injury. It is possible that such a study would reveal a more important role for the checkpoint mechanism. Our data-driven modeling approach gives novel insights on how this mechanism functions and provides biologically-motivated parameters for its function in \textit{Drosophila larva neurons}. However, we found that the checkpoint was not necessary to establish fully-biased polarity in the model presented here (though the checkpoint alone could still achieve such uniform polarity on longer timescales) and that nucleation feedback was a more efficient mechanism. We hypothesize that the checkpoint mechanism could provide robustness to noise and thus be more important in scenarios involving different initial MT distributions, with more randomness in nucleation dynamics, or for investigating polarity reversals. Additionally, while the checkpoint mechanism parameters were selected to match the biological data in this context, there may be alternative parameter regimes for other neurite systems which reflect a more influential role for the checkpoint mechanism as compared to nucleation feedback.

The mathematical framework developed here could be applied to investigate polarized filament arrays in other biological contexts. Generating polar microtubule structures in other settings can also be influenced by biased nucleation and by subsequent quality control of growth direction, independently from the underlying molecular players. For example, in axons, biased nucleation is thought to be mediated by the HAUS/augmin complex along the sides of existing microtubules \cite{Rolls2022}. This is similar to the nucleation feedback mechanism in our model, and our results suggest that this mechanism would be sufficient to maintain polarity over long time periods. Periodic nucleation sites in cellular tubes have also been proposed to generate uniformly oriented microtubules in developing astrocytes \cite{Fu2019Astrocyte}. The proposed spatial stochastic model and techniques for validation with observed polarity profiles can thus potentially extend to give insights to other biological applications where polar cytoskeleton arrangements are observed. 

\section{Acknowledgments}
This work was supported by NIH grant R01NS121245. HGS and ACN were also partially supported by NSF grant DMS-2038056. 

\appendix
\section{Supplemental figures}\label{app:supp_figures}
\begin{figure}[H]
    \centering
    \includegraphics[width=0.6\linewidth]{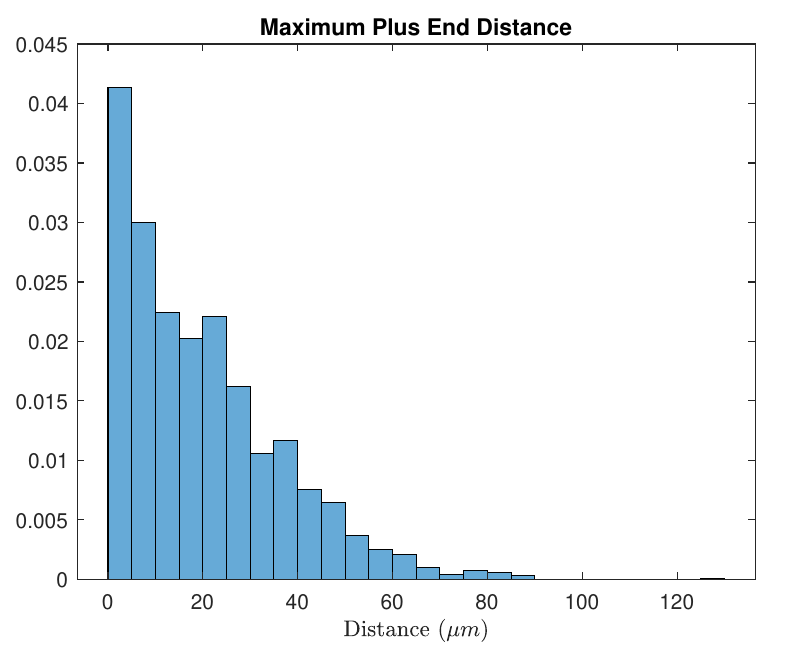}
    \caption{Distribution of the maximum distance reached by MT plus ends for 1829 simulations of full MT lifetimes, from nucleation to complete catastrophe, in the MT polymerization model. This base model does not include spatial constraints or nucleation control mechanisms. Each MT is initialized with a seed length of  $0.1~\mu$m. The fraction of MTs that grew at least $2~\mu$m is 0.9076.}
    \label{fig:plus_end_growth}
\end{figure}
\begin{figure}[H]
    \centering
    \includegraphics[width=0.8\linewidth]{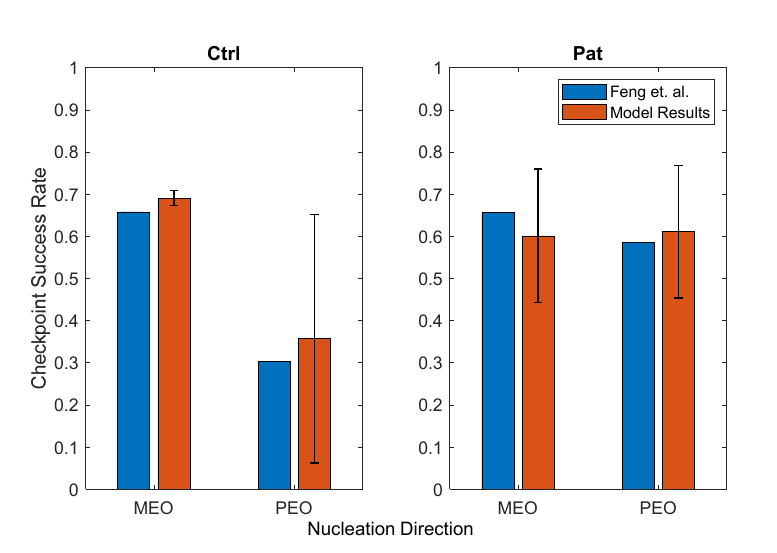}
    \caption{Model checkpoint success rates comparing results from \cite{Feng2021ParallelPolymerization} (blue) to average model results (red) with standard deviation error bars. Success rates were computed from 10 independent 30-hour simulations of the long model domain initialized with 90\% MEO and 50\% MEO MTs for the control and Patronin environments, respectively. MTs were counted as successfully passing a checkpoint if they pass the checkpoint autofail, they satisfy the matching pre-existing content condition, and they grew to be at least $2~\mu$m long. Otherwise, they were counted as failures. }
    \label{fig:checkpoint_model_success}
\end{figure}

\section{Supplemental video captions}
\label{app:supp_video_captions}

\subsection{Supplemental video 1}
\label{sec:sup_video1}
Two example simulations of our single nucleation location domain with MEO MTs in pink and PEO MTs in blue. These result in fully-biased MEO (left) and PEO (right) polarity. For both simulations, both nucleation feedback and checkpoint mechanisms were implemented. MTs are visualized from oldest to newest (based on time of nucleation) from left to right.

\subsection{Supplemental video 2} 
\label{sec:sup_video2}
Three example simulations of our long domain with 6 nucleation locations with MEO MTs in pink and PEO MTs in blue. These demonstrate spatial polarity phase separation, achieve different fully-biased polarity states, and depart fully-biased polarity states. For all simulations, both nucleation feedback and checkpoint mechanisms were implemented. MTs are visualized from oldest to newest (based on time of nucleation) from left to right.

\end{document}